\documentclass[aps, prb, amsmath, twocolumn, longbibliography, superscriptaddress, footinbib, 10pt,floatfix, final]{revtex4-2} 

\makeatletter
\def\frontmatter@maketitle{%
  \@author@finish
  \title@column\titleblock@produce
  \suppressfloats[t]%
}%
\makeatother

\usepackage{ifthen}
\usepackage{float}
\usepackage{mathtools}
\usepackage{xcolor}
\usepackage{booktabs}
\usepackage[version=4]{mhchem} 
\usepackage{physics}
\usepackage{enumitem}
\usepackage{amssymb, amsmath}
\usepackage{hyphenat}
\usepackage{comment}
\usepackage{graphicx} 
\usepackage{epstopdf}  
\usepackage{blindtext}
\usepackage{siunitx} 
\usepackage[final]{microtype} 
\usepackage{setspace}
\usepackage[T1]{fontenc}
\usepackage{lmodern}
\usepackage[colorlinks=true,linkcolor=black, citecolor=blue, urlcolor=blue]{hyperref}

\usepackage[utf8]{inputenc}
\usepackage[acronym, shortcuts]{glossaries}
\usepackage[l2tabu, orthodox]{nag}
\usepackage{makecell}
\usepackage{color,soul}
\usepackage{adjustbox}
\usepackage{multirow}
\usepackage{environ} 
\DeclareUnicodeCharacter{2212}{-}

\sethlcolor{white}

\begin{document}
\def\figureautorefname{Fig.} 
\def\tableautorefname{Table} 
\newcommand{\fref}[2]{\autoref{#1}\textcolor{blue}{#2}}

\newcommand{\paperSection}[2][normal]{%
    \ifthenelse{\equal{#1}{normal}}{%
        \medskip
        \noindent\textbf{#2}\par
    }{%
        \medskip
        \noindent\textbf{#2}\par
    }
}

\newcommand{\methodTitle}[1]{\textbf{#1}\hspace{0.5em}}

\newcommand{\GLAM}{Department of Materials Science and Engineering, Stanford University, Stanford, California 94305, USA}

\newcommand{\NIST}{Materials Measurement Science Division, Material Measurement Laboratory, National Institute of Standards and Technology, Gaithersburg, Maryland 20899, USA}

\newcommand{\MPI}{Max Planck Institute for Sustainable Materials, 40237, D{\"u}sseldorf, Germany}

\newcommand{\SLAC}{Stanford Synchrotron Radiation Lightsource (SSRL), SLAC National Accelerator Laboratory, Menlo Park, California 94025, USA}

\newcommand{\Wdc}{Department of Civil and Environmental Engineering, School of Engineering and Applied Science, George Washington University, Washington, DC 20052, USA}

\newcommand{\Berkeley}{National Center for Electron Microscopy, Molecular Foundry, Lawrence Berkeley National Laboratory, USA}

\newcommand{\BerkeleyMSE}{Department of Materials Science and Engineering, University of California Berkeley, Berkeley, California, USA }

\newcommand{\citechange}[1]{\textcolor{red}{#1}}
\newcommand{\kp}{\bm{k \cdot p}}  
\newcommand{\GeSn}[2]{\ce{Ge}$_{#1}$\ce{Sn}$_{#2}$}
\newcommand{\SiSn}[2]{\ce{Si}$_{#1}$\ce{Sn}$_{#2}$}
\newcommand{\SiGeSn}[3]{\ce{Si}$_{#1}$\ce{Ge}$_{#2}$\ce{Sn}$_{#3}$}
\newcommand{\GeSiSn}[3]{\ce{Ge}$_{#1}$\ce{Si}$_{#2}$\ce{Sn}$_{#3}$}
\newcommand{\SipGeSn}[3]{(\ce{Si})$_{#1}$\ce{Ge}$_{#2}$\ce{Sn}$_{#3}$}
\newcommand{\SiGe}[2]{\ce{Si}$_{#1}$\ce{Ge}$_{#2}$}

\newcommand{\knn}[1]{$#1$NN}
\newcommand{\quot}[1]{``#1''} 
\newcommand{\RNum}[1]{\uppercase\expandafter{\romannumeral #1\relax}}
\newcommand{\RomanNumeralCaps}[1]
    {\MakeUppercase{\romannumeral #1}}
\newenvironment{authcontrib}{\noindent \textbf{Author Contributions}}{}
\newenvironment{interests}{\noindent \textbf{Competing Interests}}{}
\newenvironment{availability}{\noindent \textbf{Data Availability}}{}


\newcommand{\citesupp}[1]{
    \footnote{See Supplemental Material at \textbf{[URL will be inserted by publisher]} for \textcolor{red}{details of the experimental setup}, which includes Refs. #1}
}


\newcommand{\figCiteChange}[1]{\colorbox{yellow!50!white}{#1}}
\DeclareSIUnit{\million}{\text{million}}


\newcommand\vertarrowbox[3][3ex]{%
  \begin{array}[t]{@{}c@{}} #2 \vspace{1ex}\\
  \left\uparrow\vcenter{\hrule height #1}\right.\kern-\nulldelimiterspace\\
  \makebox[0pt]{#3}
  \end{array}%
}

\NewEnviron{myequation}[1]{%
\begin{equation}
\scalebox{#1}{$\BODY$}
\end{equation}
}

\newcommand{\suppfref}[2]{\autoref{#1}\textcolor{blue}{#2}}

\newcommand\suppNoteRef[1]{%
    \ifthenelse{\equal{\includeSI}{1}}{%
        \let\tempsectionautorefname\sectionautorefname%
        \renewcommand\sectionautorefname{Supplementary Note}%
        \ref{#1}%
        \let\sectionautorefname\tempsectionautorefname%
    }{\textcolor{blue}{Supplementary Note \ref*{#1}}}%
}

\newcommand\suppFigRef[3]{%
    \ifthenelse{\equal{\includeSI}{1}}{%
        \let\tempfigureautorefname\figureautorefname%
        \renewcommand\figureautorefname{Supplementary Fig.}%
        \suppfref{#1}{#2}%
        \let\figureautorefname\tempfigureautorefname%
    }{\textcolor{blue}{Supplementary Fig. #3}}%
}

\newcommand\suppFigsRef[3]{%
    \ifthenelse{\equal{\includeSI}{1}}{%
        \let\tempfigureautorefname\figureautorefname%
        \renewcommand\figureautorefname{Supplementary Figs.}%
        \suppfref{#1}{#2}%
        \let\figureautorefname\tempfigureautorefname%
    }{\textcolor{blue}{Supplementary Figs. #3}}%
}

\newcommand\suppTabRef[3]{%
    \ifthenelse{\equal{\includeSI}{1}}{%
        \let\temptableautorefname\tableautorefname%
        \renewcommand\tableautorefname{Supplementary Table}%
        \suppfref{#1}{#2}%
        \let\tableautorefname\temptableautorefname%
    }{\textcolor{blue}{Supplementary Table #3}}%
}

\def\blankpage{%
      \clearpage%
      \thispagestyle{empty}%
      \null%
      \clearpage}


\epstopdfsetup{update} 
\epstopdfsetup{outdir=./figures/}
\epstopdfsetup{suffix=-generated}

\newacronym{rta}{RTA}{rapid thermal annealing}
\newacronym{ssrl}{SSRL}{Stanford Synchrotron Radiation Lightsource}
\newacronym{ald}{ALD}{atomic layer deposition}
\newacronym{exafs}{EXAFS}{extended X\hyp{}ray absorption fine structure}
\newacronym{jdos}{JDOS}{joint density of states}
\newacronym[\glslongpluralkey={nanowires}]{nw}{NW}{nanowire}
\newacronym{fwhm}{FWHM}{full\hyp{}width at half maximum}
\newacronym[\glslongpluralkey={special quasi\hyp{}random structures}]{sqs}{SQS}{special quasi\hyp{}random structure}
\newacronym{rdf}{RDF}{radial distribution function}
\newacronym{wc-sro}{WC-SRO}{Warren\hyp{}Cowley short-range order}
\newacronym[\glslongpluralkey={machine\hyp{}learning pseudopotentials}]{mlp}{MLP}{machine\hyp{}learning potential}
\newacronym{msrd}{MSRD}{mean\hyp{}square relative displacement}
\newacronym{ms}{MS}{multi\hyp{}scattering}
\newacronym{mc}{MC}{Monte\hyp{}Carlo}
 
\newacronym{vls}{VLS}{vapor\hyp{}liquid\hyp{}solid}
\newacronym{pl}{PL}{photoluminescence}
\newacronym{cvd}{CVD}{chemical vapor deposition}
\newacronym{hrxrd}{HRXRD}{high\hyp{}resolution X\hyp{}ray diffraction}
\newacronym{rpcvd}{RPCVD}{reduced\hyp{}pressure chemical vapor deposition}
\newacronym{haadf}{HAADF}{high\hyp{}angle annular dark field}
\newacronym{haadf-stem}{HAADF-STEM}{high\hyp{}angle annular dark\hyp{}field scanning transmission electron microscopy}
\newacronym{stem}{STEM}{scanning transmission electron microscopy}
\newacronym{fib}{FIB}{focus ion beam}
\newacronym{edx}{EDX}{energy\hyp{}dispersive X\hyp{}ray spectroscopy}
\newacronym{pips}{PIPS}{passivated implanted planar silicon}
\newacronym{tem}{TEM}{transmission electron microscopy}
\newacronym{hrtem}{HRTEM}{high\hyp{}resolution transmission electron microscopy}
\newacronym{fft}{FFT}{fast\hyp{}Fourier transform}
\newacronym{mct}{MCT}{mercury cadmium telluride}
\newacronym{sro}{SRO}{short\hyp{}range ordering}
\newacronym{srh}{SRH}{Shockley\hyp{}Read\hyp{}Hall}
\newacronym[\glslongpluralkey={medium\hyp{} to high\hyp{}entropy alloys}]{mhea}{M/HEA}{medium\hyp{} to high\hyp{}entropy alloy}
\newacronym{knn}{KNN}{K\hyp{}nearest neighbor}
\newacronym{sem}{SEM}{scanning electron microscopy}
\newacronym{dft}{DFT}{density functional theory}
\newacronym{vasp}{VASP}{Vienna Ab initio Simulation Package }
\newacronym{lda}{LDA}{local density approximation}

\title{Shining light on short\hyp{}range atomic ordering in semiconductors alloys}

\author{Anis Attiaoui}
\affiliation{\GLAM{}}
\affiliation{\SLAC{}}

\author{Shunda Chen}
\affiliation{\Wdc}

\author{Joseph C. Woicik}
\affiliation{\NIST{}}

\author{J. Zach Lentz}
\affiliation{\GLAM{}}


\author{Liliane M. Vogl}
\affiliation{\BerkeleyMSE{}}
\affiliation{\Berkeley{}}
\affiliation{\MPI{}}

\author{Jarod E. Meyer}
\affiliation{\GLAM{}}

\author{Kunal Mukherjee}
\affiliation{\GLAM{}}

\author{Andrew Minor}
\affiliation{\BerkeleyMSE{}}
\affiliation{\Berkeley{}}

\author{Tianshu Li}
\affiliation{\Wdc}

\author{Paul C. McIntyre}
\affiliation{\GLAM{}}
\affiliation{\SLAC{}}

\begin{abstract}
\medskip
The functional properties of semiconductors are typically controlled by tailoring their chemical composition and their state of strain, and by controlling their long-range structural order, including the presence of extended defects such as dislocations. In addition to these approaches, theoretical predictions suggest that \ac{sro} of atoms in group-\RNum{4} semiconductor alloys can modify the bandgap, a defining property of any semiconductor. Herein, a new machine learning enabled, computation-guided methodology for \ac{exafs} analysis of \ac{sro} is used to quantify the effects of local atomic order on the bandgap of germanium-tin (\GeSn{}{}) alloy single crystal nanostructures with well-controlled strain and composition. Correlative analysis of \ac{exafs} and \ac{pl} establishes the relationship between bandgap and the \ac{wc-sro} parameter of the \GeSn{}{} alloys. It is further demonstrated that \ac{sro} can be tuned over a broad range by post-deposition annealing of the alloy crystals. This work establishes control of \ac{sro} as an important design parameter for semiconducting properties and suggests the potential for quantitative measurement and tuning of \ac{sro} in other semiconductor alloy systems.

\end{abstract}
\maketitle

\par For decades, the design of crystalline semiconductor materials has rested on two pillars: controlling \textit{which atoms are present} (composition) and \textit{how the crystal is deformed} (strain). Precise control of both composition and strain are now highly reproducible approaches for material design and manufacturing underpinning the global electronics and photonics industries. Theory predicts a third pillar; controlling \textit{how atoms are arranged locally} for a given composition and strain state. \Ac{dft}~\cite{cao2020} and cluster expansion calculations~\cite{xu2019e} suggest that \acf{sro}—the preferential arrangement of chemical species within a few atomic nearest-neighbor distances—may be a fundamental and underexplored lever to tune the electronic structure of semiconductors. Similar phenomena examined theoretically in group-\RNum{4} semiconductor alloys~\cite{cao2020,chen2024}, \RNum{3}\hyp{}\RNum{5} semiconductors~\cite{bellaiche1998,liu2016}, and high-entropy alloys~\cite{singh2015} underscore the potential of \ac{sro}-driven electronic effects. However, controlled local order has remained largely inaccessible experimentally. Realizing this third pillar for practical application hinges on answering two fundamental questions. First, can \ac{sro} be manipulated deliberately and selectively to tailor semiconducting properties? Second, what approaches can enable statistically robust quantification of \ac{sro}? The absence of definitive answers to these questions has left \ac{sro} unexploited for materials design of semiconductors. A primary barrier to controlling \ac{sro} has been measurement: advanced microscopies and atom-probe based methods can map local atomic arrangements and chemical correlations~\cite{pekin2018,li2023a,he2024,vogl2025}, but statistical \ac{sro} quantification remains nontrivial. Similarly, \ac{exafs} analysis has, to date, provided qualitative insights~\cite{shimura2017,robouch2020,lentz2023a,lentz2025b,gougam2025}, but a robust quantitative framework has been lacking.

\par These challenges are addressed herein for \GeSn{}{} alloys, a material system studied for application in integrated photonics~\cite{kelzenberg2010a}, mid-infrared sensing~\cite{soref2015,moutanabbir2021b}, and quantum information~\cite{assali2022f,kaul2025}. A \ce{Ge}/\GeSn{}{} core/shell nanowire platform is employed, whereby a compliant \ce{Ge} core relaxes strain in the shell and conformal ultrathin alumina (\ce{Al2O3}) passivation enables thermal annealing high above the \ce{Ge}-\ce{Sn} eutectic temperature~\cite{braun2022a} without altering composition and strain. Thermal processing induces a pronounced, monotonic blueshift in \acf{pl}, suggesting a direct link between local atomic arrangement and optoelectronic response. To probe this link, an integrated methodology is developed, combining \ac{exafs} with high-fidelity machine-learned atomistic models within a Bayesian inference framework. This approach quantitatively extracts the \ac{wc-sro} parameter, $\alpha$, and reveals its systematic increase after annealing. Conventional mechanisms—compositional variation, strain relaxation, and defect formation—are systematically ruled out as plausible causes of the observed optical shift. The \ac{pl} blueshift, a measure of increasing bandgap, is found to be a direct consequence of increasing \ac{sro}, thus revealing a powerful approach for semiconductor materials design.

\onecolumngrid
\adjustimage{width=0.85\textwidth,center,caption=
{\textbf{Influence of \ac{sro} modulation on optoelectronic properties in \GeSn{0.9}{0.1} alloy.} \textbf{a}, The \acf{rdf} $g_{\text{Sn-Sn}}(R)$ for \knn{1} \ac{wc-sro} parameter $\alpha$, illustrating the evolution from random alloy (blue) through partial order ($\alpha = 0.5$, orange) to fully \knn{1} \ce{Sn}\hyp{}\ce{Sn} depleted ($\alpha = 1$, red) structure. The \ac{rdf} is computed from $\approx$ 1780 atoms, \acf{mlp}-generated supercells with specific \ac{sro} values. The inset shows the evolution of the ratio of the integrated intensity of the \knn{3} \ac{rdf} peak to that of the \knn{1} peak, as a function of $\alpha$. \textbf{b}, Unfolded electronic band structures back to the first Brillouin zone of the diamond cubic lattice ($L,\;\Gamma,\; \text{and}\;X$) for \acf{sqs} disordered (left) and short\hyp{}range ordered ($\alpha=1$, right) configurations, with color scale denoting the corresponding spectral weight (\hl{Methods}). The band structure is simulated at $\SI{0}{\kelvin}$. \textbf{c}, Calculated bandgap variation $\Delta E^{\text{Theo.}}_{g,\,\alpha}$ ($ =\;E_g(\alpha)-E_g(\alpha=-0.135)$) as a function of the \ac{wc-sro} parameter, for a fixed strain ($\varepsilon=\;0$) and \ce{Sn} composition ($x=10\;\text{at}.\%$). \textbf{d-e}, \Acf{sem} images of \ce{Ge}/\GeSn{}{} core/shell nanowires (NWs) in the as-grown state (blue) and after post-growth annealing at $\SI{450}{\degreeCelsius}$ (red). Both samples were coated with a $3-\SI{4}{\nm}$ layer of \ac{ald}\hyp{}\ce{Al2O3}; both scale bars: $\SI{500}{\nm}$. \textbf{f}, Normalized \acf{pl} spectra at $\SI{80}{\kelvin}$ for the as-grown (blue) and annealed (red) NWs, showing a significant post-annealing \ac{pl} blueshift ($\Delta E^{\text{Expt.}}_g$). \textbf{g}, Schematic of the Bayesian inference approach for \ac{sro} quantification, where experimental \ac{exafs} data are fit against a library of theoretical spectra generated from supercells with varying parameters $\alpha$. Iterative optimization identifies the model with the minimum R-factor, yielding the most probable ordering parameter ($\alpha^*$) for the sample.},label={fig:Fig1}, nofloat=figure, vspace=\bigskipamount}{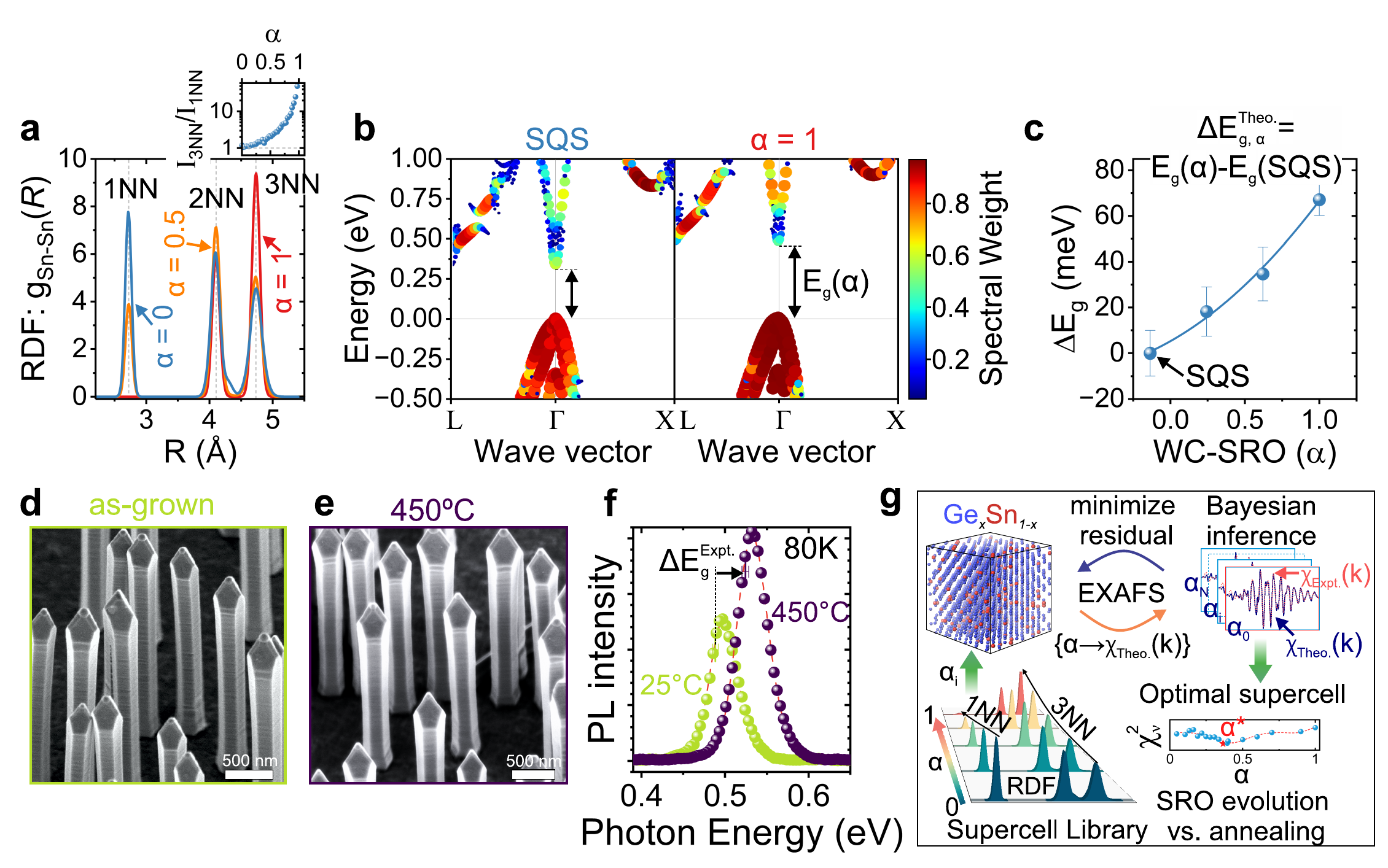}
\twocolumngrid
\paperSection{\emph{Ab initio} predictions of SRO-induced bandgap modulation}
\noindent A quantitative baseline is first constructed by predicting how conventional factors (average composition, elastic strain) affect the bandgap of \GeSn{1-x}{x} alloys (Supplementary Information section 1). This framework is then extended to incorporate short-range chemical ordering, linking the \ac{wc-sro} parameter to bandgap variations. Figure~\ref{fig:Fig1}{} summarizes theoretical insights alongside representative experimental results. Central to this framework is the $\alpha$ parameter, defined as
\begin{equation}
    \alpha_{ij}^{\text{1NN}} = 1 - \frac{P^{\text{1NN}}(i\mid j)}{c_i c_j};\quad i\neq j,
    \label{eq:1}
\end{equation}
where $P^{\text{1NN}}$ is the probability of finding $i$-$j$ atomic neighbors in the \knn{1} shell, and $c_i$ and $c_j$ are the corresponding species concentrations~\cite{cowley1950,sheriff2024}. This parameter quantifies the extent to which atomic species $i$ and $j$ preferentially cluster or avoid each other, with $\alpha \to 1$ corresponding to suppression of \ce{Sn}\hyp{}\ce{Sn} \knn{1} pairs compared to a random alloy ($\alpha = 0$). \emph{Ab initio} supercell calculations reveal evolution of the local \ac{rdf} as $\alpha$ varies: attenuation of the \knn{1} peak intensity coupled with pronounced strengthening of the \knn{3} peak (\fref{fig:Fig1}{a}).

\par Unfolded electronic structures (\fref{fig:Fig1}{b}, Methods) compare \ac{sqs} (left) and short-range ordered ($\alpha=1$, right) strain-free \GeSn{0.9}{0.1} configurations. \ac{sro} increasing from a quasi-random to a fully \knn{1} \ce{Sn}\hyp{}\ce{Sn} depleted alloy widens the direct bandgap, $E_g(\alpha)$, by $>\SI{80}{\meV}$. Figure~\ref{fig:Fig1}{c} plots the theoretical bandgap difference versus $\alpha$ at a fixed \ce{Sn} composition and zero strain, isolating the \ac{sro} effect. The predicted bandgap change rivals that induced by significant compo\hyp{}
\onecolumngrid
\adjustimage{width=.75\textwidth,center,caption=[\protect\suppFigRef{fig:figSI_HRTEM_All}{}{}]
{\textbf{Structural and compositional characterization of \ce{Ge}/\GeSn{}{} core/shell nanowires (\glspl{nw}) after rapid thermal annealing.} \textbf{a–d}, \ac{sem} micrographs of NW arrays post-growth annealing ($\SIrange{300}{450}{\degreeCelsius}$). \textbf{e}, High\hyp{}resolution X\hyp{}ray diffraction $2\theta$–$\omega$ scans around the $(333)$ reflection. The intense peak at $\SI{90.05}{\degree}$ corresponds to the \ce{Ge} $(333)$ substrate; the broader peak at $\SI{88.5}{\degree}$ arises from the \GeSn{}{} shell. \ce{Sn} content determination is described in Methods. Solid lines: experimental data from annealed samples. Dotted lines: experimental data from as-grown samples. All samples were coated with a $\approx \SI{3}{\nm}$ alumina layer. \textbf{f}, Side\hyp{}view \ac{tem} of as-grown reference \ce{Ge}/\GeSn{}{} nanowire structure. \textbf{g}, Side\hyp{}view \ac{tem} of representative annealed nanowire ($\SI{450}{\degreeCelsius}$), with corresponding \ac{edx} elemental maps displaying the spatial distributions of \ce{Ge} and \ce{Sn}. \textbf{h}, Cross-sectional \ac{edx} color overlay map. \textbf{i}, \ac{haadf-stem} image of a single NW annealed at $\SI{450}{\degreeCelsius}$, with indexed $\{112\}$ and $\{110\}$ crystal facets. \textbf{j}, \ac{hrtem} image of the \ce{Ge}/\GeSn{}{} interface. Inset (upper left): diffraction pattern image indexed to the $[111]$ zone axis. Inset (lower right): aberration-corrected lattice image at the heterointerface. \textbf{k}, \ac{hrtem} image at the \GeSn{}{}/\ce{Al2O3} interface with an inset showing an aberration-corrected lattice image at the interface.},label={fig:Fig2},nofloat=figure,vspace=\bigskipamount}{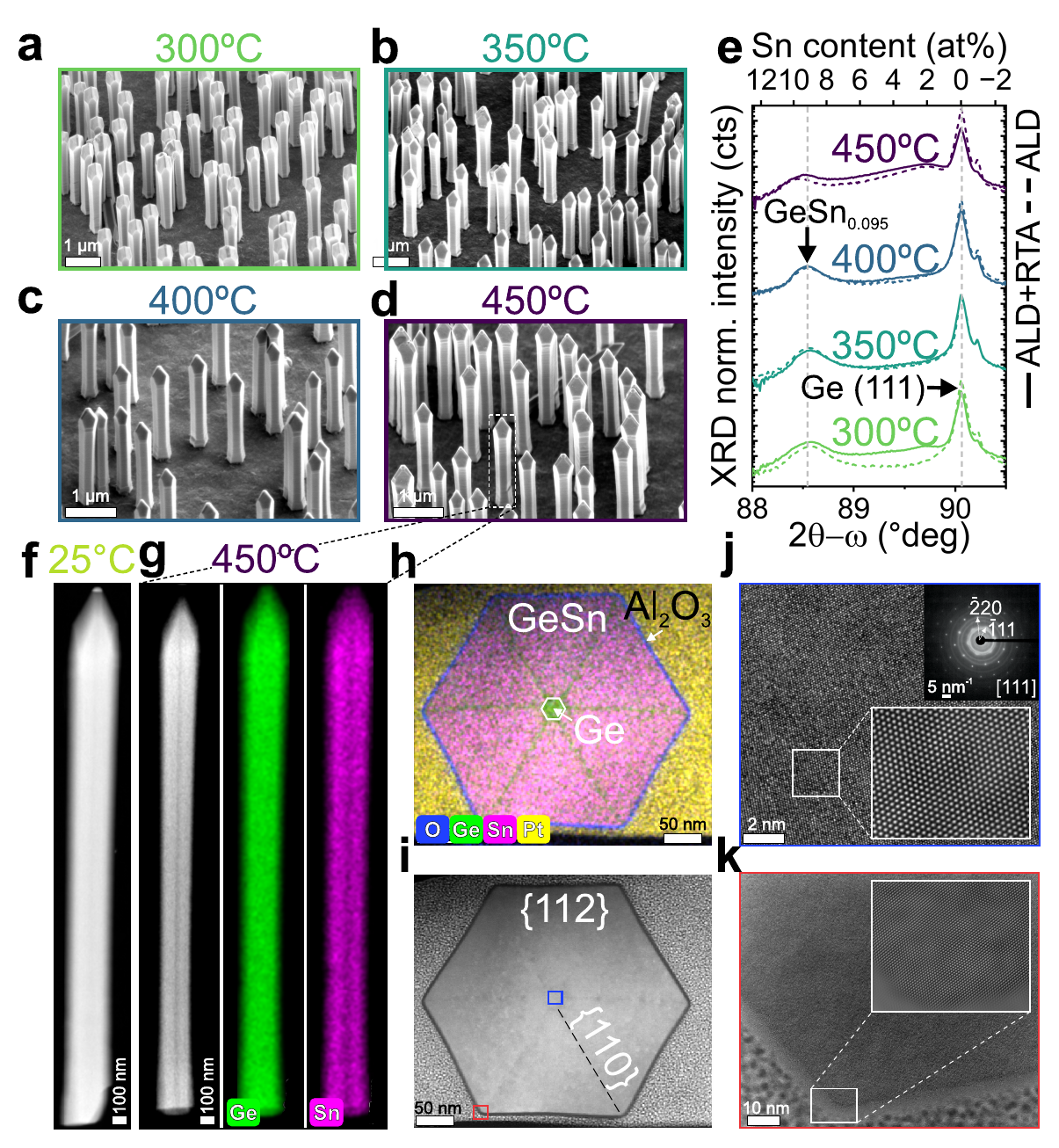}
\twocolumngrid
\begin{table*}[!ht]
  \begin{center}
  \caption{\textbf{Structural, compositional, and electronic properties of post-annealed \ce{Ge}/\GeSn{}{} core/shell nanowires.} \ce{Sn} content is measured by \ac{hrxrd} and \ac{tem}\hyp{}\ac{edx}, thickness by \ac{sem}, and bandgap shift $\Delta E_g^{\mathrm{Expt.}} = E_g(T) - E_g(25^{\circ}\mathrm{C})$ in meV by infrared PL spectroscopy at $80~\mathrm{K}$. $T$ denotes the post-growth annealing temperature. Theoretical bandgap shifts $\Delta E_{g,\,x}^{\mathrm{Theo.}}$ (meV) is computed via \ac{dft} using \ac{sqs} supercells that represent the measured alloy disorder at the measured \ce{Sn} content. $\Delta E_{g,\,x}^{\mathrm{Theo.}}$ isolates composition-driven changes at fixed strain ($\varepsilon=0$) by varying \ce{Sn} content ($x$) in \ac{sqs} cells. 
  Finally, for each entry, the \ac{wc-sro} parameter is reported as $\alpha \,(\pm\,\sigma_{\alpha})$, inferred via the adaptive Bayesian \ac{exafs} fitting framework that integrates \ac{dft}-generated supercells. The average \ac{wc-sro} uncertainty $(\sigma_{\alpha})$ is estimated to be $0.05$. Reported uncertainties reflect the measurements or computation associated with each technique/estimate.}
  \label{tab:Table1}
  \vspace{2ex}
  \setlength{\tabcolsep}{4pt}
  \renewcommand{\arraystretch}{1.15}
  \begin{tabular}{@{}cccccccc@{}}
    \toprule\midrule
    \textbf{Samples} & 
    \makecell[c]{\textbf{$\boldsymbol{T}$} \\ \textbf{($\pm\;5^{\circ}\text{C}$)}} & 
    \makecell[c]{\textbf{$\boldsymbol{d_{\mathrm{shell}}}$} \\ \textbf{(nm)}} & 
    \makecell[c]{\textbf{Sn content} \\ \textbf{($\pm\;0.5$ at. \%)}\textcolor{blue}{\textsuperscript{a}}}& 
    \makecell[c]{\textbf{$\boldsymbol{E^{RT}_{g}\;\rightarrow\;E^{T}_{g}}$} \\ \textbf{($\pm\;\SI{5}{\meV}$)}} &  
    \makecell[c]{\textbf{$\boldsymbol{\Delta E^{\text{Expt.}}_{g}}$} \\ \textbf{($\pm\;\SI{5}{\meV}$)}} & 
    \makecell[c]{\textbf{$\boldsymbol{\Delta E_{g,\,x}^{\text{Theo.}}}$} \\ \textbf{(meV)} \\ \textbf{(DFT)}} & \makecell[c]{\textbf{SRO} \\ \textbf{($\boldsymbol{\alpha}$)}\\ ($\pm0.05$)} \\ 
    \midrule
    A & N.A. & $140\pm7$  & $9.5$ & $481\rightarrow\text{N.A.}$ & 0 & \text{N.A.} & 0.20 \\
    B & 300 & $145\pm7$  & $9.5$ & $481\rightarrow490$  & 8 & 4 & 0.30 \\
    C & 350 & $146\pm10$  & $9.5$ & $489\rightarrow502$  & 12 & 5 & 0.36 \\
    D & 400 & $141\pm7$  & $9.5$ & $486\rightarrow509$  & 21 & 7 & 0.43 \\
    E & 450 & $139\pm12$  & $9.6$ & $489\rightarrow515$  & 25 & 8 & 0.52 \\
    \midrule\bottomrule
  \end{tabular}
  \end{center}

  \vspace{1ex}

  \parbox{\linewidth}{%
  \footnotesize
  \textcolor{blue}{\textsuperscript{a}} Rounded to nearest 0.1 at.\%; sample-to-sample variations ($\leq0.1$ at.\%) are below measurement precision.\\
  }
\end{table*}
\noindent sitional or strain variations~\cite{gupta2013}. For example, achieving $\SI{10}{\meV}$ bandgap shift via composition change requires $\approx0.75\;\text{at.}\%$ \ce{Sn} reduction. For intermediate \ac{sro} values, supercells are generated using \ac{mlp} (Methods), enabling efficient statistical estimation of the bandgap across a wide range of \ac{wc-sro} values ($0<\alpha\leq1$). Random alloys exhibit lattice disorder-induced electronic state localization, narrowing the bandgap. Conversely, greater \ac{sro} fosters a more regular atomic motif, reduces local strain, and promotes electronic delocalization, collectively increasing the bandgap~\cite{liang2024}.

\par Experimental evidence supporting \ac{sro}-driven bandgap modulation is highlighted in \fref{fig:Fig1}{d-f}. Panels d and e display \ac{sem} images of as-grown and annealed \ce{Ge}/\GeSn{}{} nanowire ensembles, demonstrating morphological stability. Panel f shows normalized infrared \ac{pl} spectra at $\SI{80}{\kelvin}$: annealed samples exhibit significant blueshift ($\Delta E^{\text{Expt.}}_{g,\,\alpha}$). To link optoelectronic changes to atomic-scale structure, an adaptive Bayesian inference algorithm resolves the $\alpha$ parameter directly from \ac{exafs} spectra. The workflow, illustrated in \fref{fig:Fig1}{g} (Supplementary  Information Figs. 4-5), links experimental data with first-principles structural models. This model-based approach moves beyond conventional \ac{exafs} analysis (Supplementary Information section 2) by mapping experimental spectra into a physically meaningful order parameter, enabling statistically robust \ac{sro} quantification.

\paperSection{Sample preparation and multimodal characterization}
\noindent Single-crystalline \ce{Ge}/\GeSn{}{} core/shell nanowires were epitaxially grown on \ce{Ge}~$(111)$ substrates (Methods). \Ac{sem} verified vertical alignment, uniform diameter, and morphological stability (\fref{fig:Fig2}{a–d}). \Ac{hrxrd} validated phase purity, $\langle111\rangle$ crystallographic orientation, and yielded an ensemble-averaged nominal \ce{Sn} content of $(9.50 \pm 0.25)\;\text{at.}\%$ (\fref{fig:Fig2}{e}). A reference plan-view \ac{tem} image of the as-grown structure is shown in \fref{fig:Fig2}{f}. \Ac{edx} in the \ac{tem} mapped elemental distributions (\fref{fig:Fig2}{g-h}), confirming uniform \ce{Sn} content of $(9.40 \pm 0.75)\;\text{at.}\%$ (Methods, Supplementary  Information Fig.~1 and \fref{tab:Table1}{}).

\par A central challenge in tuning \ac{sro} in \GeSn{}{} alloys is their intrinsic metastability and strong tendency for \ce{Sn} segregation above the eutectic point~\cite{chen2013}. \Ac{rta} is employed on nanowire assemblies encapsulated by an ultrathin ($\approx3$--$\SI{4}{\nm}$), conformal alumina layer deposited via \acf{ald} (Methods). This capping inhibits \ce{Sn} surface segregation, a primary kinetic pathway for \GeSn{}{} partial melting~\cite{braun2022a}, while preserving crystal quality and optical transparency~\cite{berghuis2021} during annealing. Studied annealing temperatures spanning $\SIrange{300}{450}{\degreeCelsius}$ may permit atomic diffusion to reorganize local bonding within the \GeSn{}{} alloy shells.

\paperSection{Post-annealing nanowire luminescence}
\noindent Low-temperature infrared \ac{pl} spectroscopy elucidates the influence of \ac{rta} on the optical response of \GeSn{}{} \glspl{nw} (Methods). The core/shell heterostructure shapes emission characteristics through three mechanisms. First, at compositions exceeding $\approx8\;\text{at.}\%$ \ce{Sn}, the shell undergoes an indirect-to-direct bandgap transition~\cite{Attiaoui2014}, rendering $\Gamma$-valley emission dominant. The resulting type-\RNum{1} band alignment confines both electrons and holes to the \GeSn{}{} shell. Second, when shell thickness ($\approx\SI{150}{\nm}$) exceeds optical penetration depth ($\approx\SI{100}{\nm}$ at $\SI{808}{\nm}$ laser excitation~\cite{Assali2021a}), photoexcitation is largely shell-confined, improving radiative efficiency and enabling direct measurement of bandgap modulation. Third, strain partitioning between core and shell enables wavelength engineering~\cite{meng2019a,assali2017a}. Consequently, \ac{pl} spectra reflect shell-dominated, direct-gap recombination, establishing a baseline for examining \ac{rta}-induced modifications.

\par Figure~\ref{fig:Fig3}{a} displays normalized \ac{pl} spectra acquired at $\SI{80}{\kelvin}$ for samples annealed at selected temperatures (\fref{tab:Table1}{}), benchmarked against unannealed alumina-coated controls. Alumina encapsulation preserves spectral characteristics while providing modest ($\approx10\%$) intensity gain, possibly due to suppressed non-radiative recombination via surface passivation of dangling bonds~\cite{gupta2013b}. Three trends emerge across all annealed samples: (i) a monotonic blueshift commensurate with rising annealing temperature, indicating progressive bandgap widening; (ii) pronounced \ac{pl} intensity enhancement, reaching twenty-fold at $\SI{450}{\degreeCelsius}$, consistent with reduced non-radiative recombination; and (iii) stable \ac{fwhm}, suggesting uniform average composition. Spectra exhibit single symmetric peaks (Supplementary  Information Fig.~7, Supplementary Information section 3) confirming that the $\Gamma$-minimum lies sufficiently below the 
\onecolumngrid
\adjustimage{width=0.73\textwidth,center,caption=
{\textbf{Spectroscopic and structural evolution of \ce{Ge}/\GeSn{}{} with thermal annealing.} \textbf{a}, $\SI{80}{\kelvin}$ \ac{pl} spectra for nanowires annealed across a range of temperatures ($\SIrange{300}{450}{\degreeCelsius}$) compared to as-grown, unannealed ($\SI{25}{\degreeCelsius}$), alumina-coated control samples (yellow-green curves); all are normalized to their peak maximum. The graded colored band illustrates the bandgap blueshift with $T$. Dashed red lines denote the \ac{jdos} model line shape fits (\fref{eq:jdos_lineshape}{} in Methods). The excitation power density was fixed to $\SI{1.9}{\kilo\watt\per\cm\squared}$ for all samples. \textbf{b}, Extracted bandgap shift $\Delta E^{\text{Expt.}}_g$ plotted against annealing temperature (dashed line is guide to the eye). Error bars denote three standard deviations. \textbf{c}, $k^2$-weighted \ac{exafs} spectra, $k^2\cdot\chi(k)$, acquired at the \ce{Sn}-K edge. The line plotted below each spectrum displays the normalized residual. For all temperatures, the residuals remain below $10^{-2}$. \textbf{d}, Corresponding Fourier transforms, $|\mathrm{FT}(k^2\cdot\chi(k))|$ as a function of radial distance $R$. Each trace is segmented into two regions: the unshaded area (left) highlights the dominant \knn{1} peak with its associated y-axis, and the shaded gray region (right) isolates the typically weaker \knn{2} and \knn{3} peaks, for which a secondary (right-side) y-axis with expanded scale is provided. Dashed gray lines serve as guides to the eye for each peak.},label={fig:Fig3},nofloat=figure,vspace=\bigskipamount}{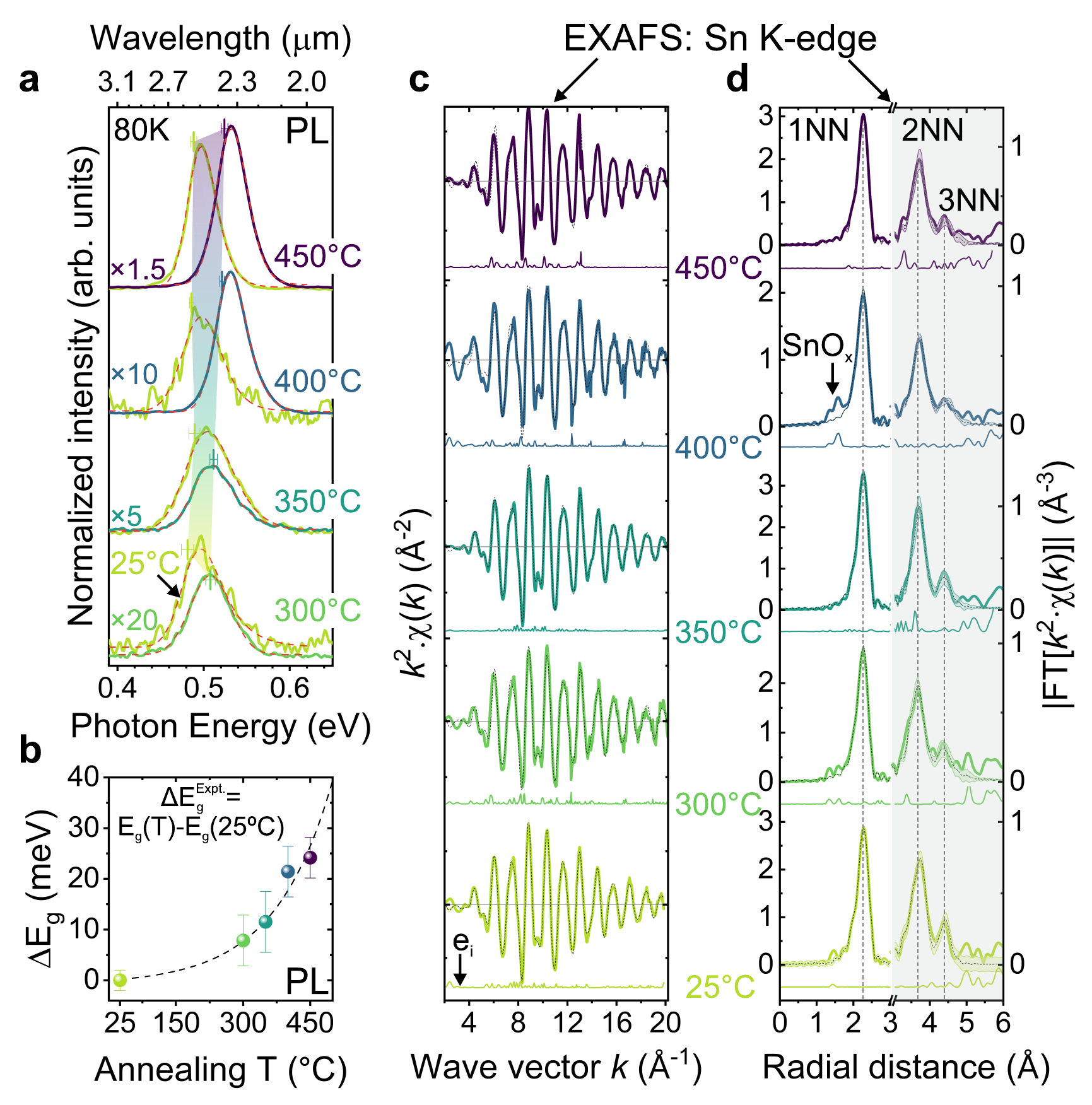}
\twocolumngrid
\noindent $L$-minimum to suppress intervalley tunneling~\cite{pezzoli2014}. The as-grown reference exhibits a narrow average \ac{pl} bandgap of $488 \pm \SI{4}{\meV}$, consistent with \ac{tem}-\ac{edx} and \ac{hrxrd} evidence of shell compositional stability and closely aligned with prior literature~\cite{assali2017a} for nanowires of similar geometry and composition (Supplementary Information section 1).

\par Temperature-dependent \ac{pl} emission from the annealed samples (Supplementary  Information Fig.~7) shows systematic shifts and broadening from $\SI{80}{\kelvin}$ to $\SI{300}{\kelvin}$, consistent with band-to-band recombination and intrinsic thermal effects. The pronounced intensity rise upon cooling is a defining characteristic of direct bandgap semiconductors~\cite{pavesi1994}, corroborated by fits to the combined \ac{jdos}/Vi\~{n}a model~\cite{vina1984c} (Supplementary Information section 3, Methods). Figure~\ref{fig:Fig3}{b} summarizes the experimental bandgap shift versus annealing temperature. The measured bandgap closely mirrors theoretical predictions (\fref{fig:Fig1}{c}), suggesting that enhanced chemical \ac{sro} drives the observed \ac{pl} blueshift. However, identifying the blueshift origin requires a critical evaluation of plausible contributing factors other than \ac{sro}.

\paperSection{Plausible annealing-induced blueshift mechanisms}
\noindent Four plausible mechanisms for the observed spectral shift during annealing are evaluated: \ce{Sn} composition reduction in the \GeSn{}{} shell, a change in its elastic strain state, crystalline defect formation, and increased \ac{sro}.

\noindent\textit{\ce{Sn} composition:} Bandgap shifts may arise from changes in average \ce{Sn} content, as the \GeSn{}{} bandgap increases with decreasing \ce{Sn} concentration~\cite{driesch2015}. To ascertain whether post-growth annealing modifies shell composition, \ac{hrxrd} and \ac{tem}-\ac{edx} measurements were compared across the annealed sample series. Both techniques confirm stable \ce{Sn} content of $(9.5 \pm 0.5)\;\text{at.}\%$ independent of thermal history (\fref{fig:Fig2}{e, g-h}, \fref{tab:Table1}{}) with no systematic trends correlating with annealing temperature. \Ac{edx} maps (\fref{fig:Fig2}{e,f}; Supplementary Information Fig.~S1) show uniform \ce{Sn} distribution with no evidence of clustering, segregation, or secondary phase formation. The characteristic faceted \quot{sunburst} \GeSn{}{} shell morphology remains unaltered~\cite{assali2017a}.

\par To quantify how subtle compositional variations would alter the bandgap, first-principles \ac{dft} sensitivity simulations were conducted (zero strain, $\alpha=0$), spanning \ce{Sn} compositions from $9-11\;\text{at.}\%$ (Supplementary Information section 1). The calculations predict a maximum sensitivity of $\approx\SI{35}{\meV}/\text{at.}\%$ \ce{Sn}, in agreement with experimental reports of $\approx30-\SI{40}{\meV}\;\text{at.}\%$ \ce{Sn} for \GeSn{}{} alloys~\cite{gallagher2014}. Assuming maximum compositional uncertainty ($\pm1.0\;\text{at.}\%$, combining \ac{hrxrd} and \ac{edx} errors), composition-driven bandgap shifts should not exceed $\SI{35}{\meV}$, nor exhibit a systematic monotonic increase with annealing temperature (\fref{fig:Fig3}{b}). The absence of detectable \ce{Sn} redistribution, which is likely required for a notable change in average shell composition, also indicates this mechanism is not the cause of the measured \ac{pl} blueshift.\smallskip

\noindent\textit{Strain state:} Lattice mismatch between the \ce{Ge} core and \GeSn{0.905}{0.095} shell may induce residual strain, thereby altering the bandgap~\cite{Wirths2015a,Attiaoui2014}. However, elasticity theory~\cite{wang2019a} predicts that coherent core/shell nanowires achieve $>\SI{95}{\percent}$ strain relaxation in the shell when the shell thickness-to-core radius ratio exceeds $3$. For the geometry studied herein (ratio $\approx5.6$), the \GeSn{}{} shell is predicted to be nearly fully strain relaxed. Moreover, \ac{hrxrd} $\omega-2\theta$ scans of the symmetric $(333)$ Bragg reflection (\fref{fig:Fig2}{e}) confirm this: the \GeSn{}{} shell peak position remains invariant, indicating negligible lattice parameter change during annealing. If residual compressive strain in \GeSn{}{} were relaxing, the bandgap should decrease~\cite{Wirths2015a}. The monotonic blueshift (\fref{fig:Fig3}{b}) is, therefore, inconsistent with strain-driven bandgap modulation.

\par Aberration-corrected \ac{haadf-stem} and \ac{hrtem} (\fref{fig:Fig2}{i–k}, Supplementary Information Fig.~S1) corroborate this, revealing defect-free \ce{Ge}/\ce{GeSn} epitaxy and an abrupt \ce{GeSn}/\ce{Al2O3} interface. Diffraction patterns from the core region (\fref{fig:Fig2}{j}, top inset) indicate single-crystalline structure is maintained after annealing. The agreement between elasticity theory predictions, \ac{hrxrd} lattice parameter stability, and atomic-resolution imaging confirms both composition and strain state remain effectively constant, eliminating these as plausible mechanisms for the \GeSn{}{} bandgap change.\smallskip

\noindent\textit{Shell crystalline imperfections:} Annealing could, in principle, introduce crystalline defects (e.g., dislocations, stacking faults, impurity clusters) that degrade optical performance. TEM performed across the annealed series consistently reveal no evidence of such defects (\fref{fig:Fig2}{i-k}, Supplementary Information Fig.~S1). The power-law dependence of \ac{pl} intensity on excitation power density ($I_{\text{PL}}\propto P^{m}$, Supplementary Information section 3) provides insight into defect-related carrier recombination mechanisms. For the as-grown sample, linear scaling ($m=1.04 \pm 0.02$) reflects low-injection operation, where background carrier concentration (associated with unintentional doping~\cite{seifner2019a}) exceeds photo-generated carriers across the measured power range. Annealing reduces background carriers via defect passivation, enabling partial access to the high-injection regime at higher powers. The higher exponent ($m=1.76 \pm 0.05$), approaching the $m\approx2$ limit characteristic of \ac{srh} recombination, is consistent with the observed twenty-fold \ac{pl} intensity amplification (\fref{fig:Fig3}{a}): defect passivation and/or annihilation reduces the non-radiative loss rate, improving the internal quantum efficiency~\cite{gerber2017}. These observations indicate that, although defect populations are modified during annealing, they do not account for the observed blueshift.\smallskip

\noindent\textit{\ac{sro} evolution:} We now consider whether \ac{rta} measurably modifies \ac{sro}. \Ac{exafs} measurements at the \ce{Sn} and \ce{Ge} K-edges (\fref{fig:Fig3}{c-d}; Supplementary Information Fig.~S9) are combined with first principles–based supercell modeling and adaptive Bayesian inference to extract values of the \ac{wc-sro} parameter $\alpha$ across the annealed sample series. Adaptive Bayesian inference identifies, for each dataset, the most probable \ac{sro} state from a library of \ac{dft}-based \ac{mlp} generated supercells and returns credible intervals of \ac{exafs}-related parameters (Supplementary Information section 4).

\par The Fourier transform magnitudes, $|\mathrm{FT}(k^{w}\cdot\chi(k))|$, exhibit peak positions tracking mean neighbor distances and widths reflecting the \ac{msrd}~\cite{timoshenko2021}. Nonetheless, phase shifts, windowing/$k$-weight effects, and multiple scattering limit conventional analysis to qualitative interpretation (Supplementary Information section 2). The measured \ac{exafs} spectra evolve systematically with annealing (\fref{fig:Fig3}{d}). Chemically selective single-scattering path decomposition from both K-edges (\fref{fig:Fig4}{a}, Supplementary Information Fig.~S9; Supplementary Information section 5) decouples \ce{Sn}$^*$\hyp{}\ce{Sn} from \ce{Ge}$^*$\hyp{}\ce{Sn} coordination. The \ce{Ge}$^*$\hyp{}\ce{Sn} scattering path intensities increase consistently across coordination shells with rising annealing temperature, indicating enhanced \ce{Ge}\hyp{}\ce{Sn} bonding (\fref{fig:Fig4}{a}). This trend is mirrored in the \ce{Sn}$^*$\hyp{}\ce{Ge} scattering paths (Supplementary Information Fig.~S9), consistent with depletion of \knn{1} \ce{Sn}\hyp{}\ce{Sn} pairs and with \ac{rdf} calculations (Supplementary Information Fig.~S2).

\par The \ce{Sn}$^*$\hyp{}\ce{Sn} paths reveal a more complex evolution. While the \knn{1} peak intensity decreases monotonically with annealing, the \knn{3} peak enhancement saturates or weakly reverses at the highest annealing temperature. This non-monotonic \knn{3} behavior is consistent with 
\onecolumngrid
\adjustimage{width=0.75\textwidth,center,caption=
{\textbf{Bayesian inference short-range order quantification and bandgap-SRO correlation.} \textbf{a}, Temperature-dependent single-scattering path analysis for first three coordination shells (\knn{1}, \knn{2}, \knn{3}) \ce{Sn}$^*$–\ce{Sn} (top) and \ce{Ge}$^*$–\ce{Sn} (bottom) pairs. Asterisk denotes absorber atom. Aggregate \ac{exafs} contributions from these paths, isolate element-specific local coordination environments. \textbf{b}, The \ac{msrd} ($\sigma^2$) for \ce{Sn}$^*$–\ce{Sn} pairs across coordination shells. Colored bars: theoretical \ac{msrd} from \acf{mlp}\hyp{}supercells ($\SIrange{25}{450}{\degreeCelsius}$); bar heights reflect ensemble variance at fixed $\alpha$. Circles: experimental \ac{msrd} extracted from \ac{exafs} Bayesian inference; error bars indicate posterior uncertainties (Methods). \textit{Insets}: \acf{wc-sro} parameter $\alpha$ versus annealing temperature (circles, right). Integrated intensity ratio of \knn{3} to \knn{1} ($I_{\text{3NN}}/I_{\text{1NN}}$) of \ce{Sn}$^*$–\ce{Sn} peaks in panel a (stars, left). \textbf{c}, Bayesian inference validation through \ac{msrd}: experimental ($\sigma_{\text{Expt.}}^2$) versus theoretical ($\sigma_{\text{Theo.}}^2$) for \ce{Sn}$^*$–\ce{Sn} scattering pairs across \knn{1} and \knn{3} coordination shells ($\SIrange{25}{450}{\degreeCelsius}$). Colored ellipses represent $3\sigma$ (95\;\% confidence) centroid estimation of the data distribution for each shell. \textbf{d}, Bandgap shifts vs. the \ac{wc-sro} parameter $\alpha$: experimental ($\Delta E_g^{\text{Expt.}}$, from \ac{pl}) and theoretical ($\Delta E_{g,\,\alpha}^{\text{Theo.}}$, from \ac{dft}) relative to the as-grown sample \ac{wc-sro} parameter ($\alpha_{25^{\circ}\text{C}}=0.20\pm0.05$) quantified using the current Bayesian inference. Error bars: vertical (from structural and optical measurements); horizontal (from statistical uncertainty in $\alpha$ from Bayesian analysis). Red curve and band show \ac{dft} ensemble mean and variance.},label={fig:Fig4}, nofloat=figure, vspace=\bigskipamount}{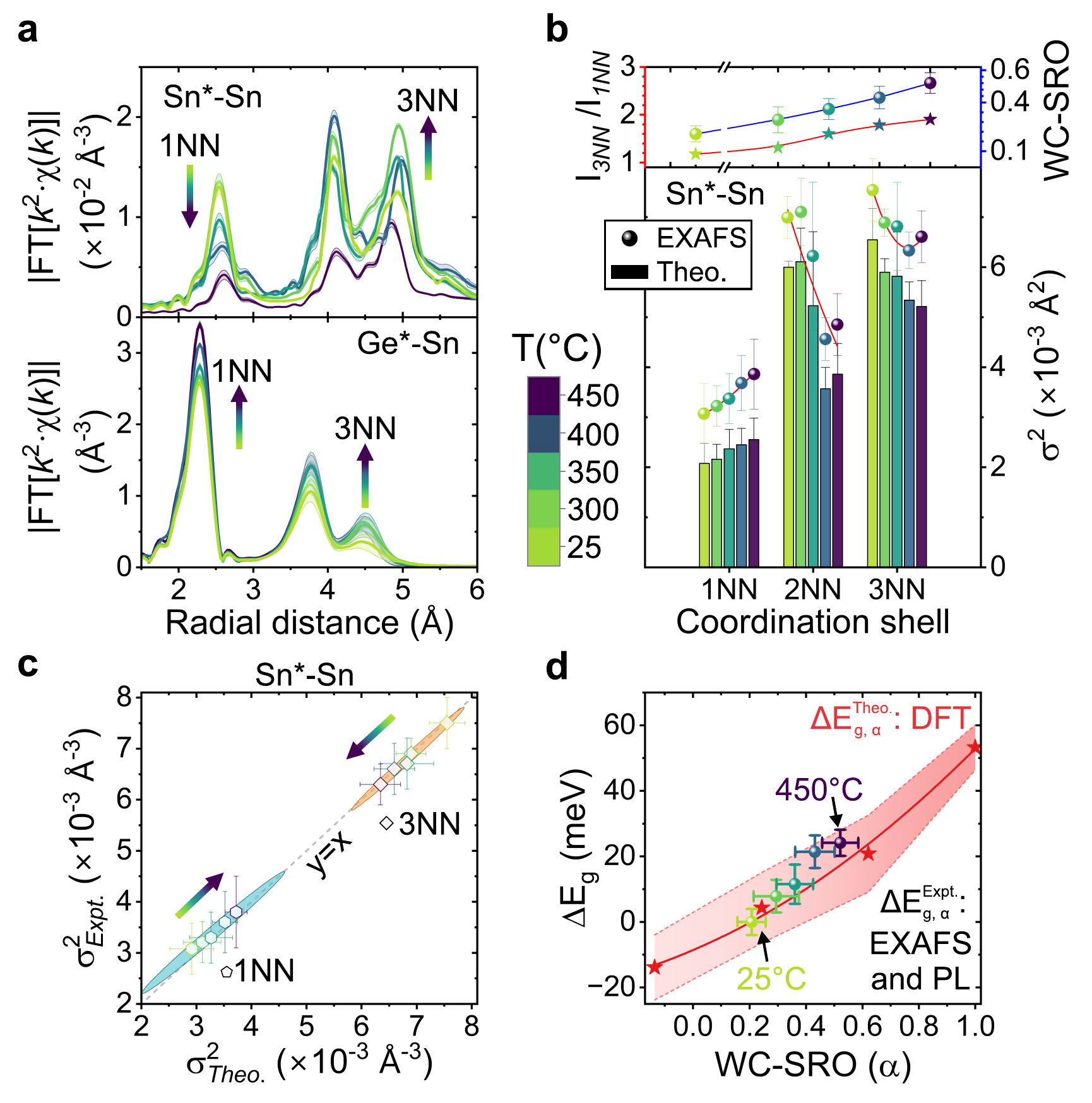}
\twocolumngrid
\noindent \ac{ms} interference effects in partially ordered crystalline systems, where changes in the chemical identity of intermediate scattering atoms shift \ac{ms} path phases~\cite{rehr2000}. The Bayesian framework, by fitting the entire experimental \ac{exafs} spectrum, implicitly accounts for \ac{ms} interference to quantify the \ac{sro} order parameter, $\alpha$, versus annealing temperature (\fref{fig:Fig4}{b}).

\par Within the Bayesian inference framework, each best-fit supercell carries a distinct \ac{wc-sro} value, enabling mapping between \ac{exafs} features and $\alpha$. The \ac{msrd} is computed independently for each scattering path and coordination shell (Supplementary Information section 2)~\cite{woicik2023}. The bond lengths are also evaluated (Supplementary Information Table~1). The \ac{wc-sro} parameter $\alpha$ increases monotonically from $0.20\;(\pm0.05)$ for the as-grown material to $0.52\;(\pm0.05)$ after annealing at $\SI{450}{\degreeCelsius}$ (\fref{fig:Fig4}{b}, top inset, left). Progressive redistribution of \ce{Sn} atoms from first- to third-neighbor shells is tracked by the \knn{3}-to-\knn{1} peak integrated intensity ratio (\fref{fig:Fig4}{b}, top inset, right), mirroring the predicted evolution of the \ac{rdf} (\fref{fig:Fig1}{a}, Extended Data Fig.~6d) and providing a self-consistent link between theory and experiment. Shell-resolved \acp{msrd} (\fref{fig:Fig4}{b}) quantify the atomistic reordering. While the extracted values are consistent with prior reports~\cite{lentz2023a,gougam2025}, the adaptive Bayesian framework greatly reduces statistical uncertainty (Supplementary Information Table~1, Supplementary Information section 4).

\paperSection{Discussion and Outlook}
\noindent An increase in \GeSn{}{} bandgap after annealing is attributed to increasing \ac{sro} resulting from local atomic redistribution. This conclusion is reinforced by considering three independent observables: (1) local structural order inferred from \ac{exafs}, (2) bandgap shifts from \ac{pl}, and (3) first-principles ensemble predictions. Figure~\ref{fig:Fig4}{c} compares experimental and theoretical \ac{msrd} for \ce{Sn}$^*$–\ce{Sn} scattering pairs at the inferred $\alpha$ for each sample (\fref{tab:Table1}{}). Strong correlation around the $y = x$ line for the \knn{1} and \knn{3} coordination shells across the annealing series demonstrates internal consistency, indicating that the model-based inference successfully captures the local  order probed by \ac{exafs}. The overlaid first-principles predictions from 128-atom supercell ensembles in Figure~\ref{fig:Fig4}{d} are not fitted data; rather, $\alpha$ values are independently extracted from \ac{exafs}, and the supercells are constructed at those $\alpha$ values. The red curve traces the ensemble mean of calculated bandgap shifts, and the shaded band reports configurational variability. For partially ordered alloys, different local atomic arrangements can yield identical mean $\alpha$ but distinct bandgaps due to motif-level heterogeneity (Supplementary Information section 6).

\par A bandgap–\ac{sro} sensitivity metric $S=\mathrm{d}E_g/\mathrm{d}\alpha$ with an experimental slope $79 \pm 10\;\SI{}{\meV}/\alpha$ agrees with the \ac{dft}–ensemble trend ($65 \pm 6\;\SI{}{\meV}/\alpha$) within uncertainty. This agreement is consistent with \ac{sro} being the principal driver of observed bandgap changes across the explored annealing range. Atomic-level analysis pinpoints local \ce{Sn}–\ce{Sn} motif heterogeneity as the source of residual variation: larger bandgaps correlate with suppressed nearest-neighbor \ce{Sn}\hyp{}\ce{Sn} pairs and an augmented third-neighbor population. The mean \ac{sro} is, therefore, an effective descriptor of the trend, but the distribution of local motifs contributes to the bandgap variance at fixed \ac{sro}~\cite{jin2023}.

\par The ability to both quantify the extent of \ac{sro} and predict its effect on the bandgap enables a significant advance in controlling this overlooked degree of freedom for semiconductor alloy design. Intentional tuning of \ac{sro} may be broadly useful across multiple classes of such materials. Specific to the \GeSn{}{} system, this work establishes thermal processing as a viable approach for manipulating atomic short-range order, thus engineering the bandgap independent of changes in composition or strain of the crystal.


\bigskip \medskip
\paperSection{METHODS}
\methodTitle{First-Principles Calculations.} Structural models of \GeSn{}{} alloys containing $1728$ atoms with 10~at.\% \ce{Sn} with varying degrees of \ac{sro} were generated via Monte Carlo sampling at a range of temperatures (\SIrange{300}{1500}{\kelvin} with \SI{300}{\kelvin} step) using state\hyp{}of\hyp{}the\hyp{}art and highly accurate \ac{mlp} for \GeSn{}{} alloys~\cite{chen2024}. For band structure calculations (\fref{fig:Fig1}{b}), all $128$\hyp{}atoms generated structures with variable \ac{sro} were further optimized with \ac{dft} using local \ac{lda} for the exchange-correlation functional, known for yielding excellent agreement with experimental results for geometry optimization in pure \ce{Ge} and \ce{Sn}~\cite{cao2020, jin2022,jin2023, chen2024}. A Monkhorst-Pack $2\times2\times2$ $k$-point mesh~\cite{Monkhorst1976} and a plane-wave cutoff energy of $\SI{300}{\eV}$ were applied. Structural relaxations during total energy minimization were performed with the conjugate gradient algorithm, imposing convergence criteria of $\SI{1e-4}{\eV}$  and $\SI{1e-3}{\eV}$ for electronic and ionic steps, respectively. To improve the fidelity of bandgap estimations, the modified Becke-Johnson (mBJ) exchange potential~\cite{tran2009} combined with \ac{lda} correlation was employed, as implemented in the \ac{vasp} software package~\cite{kresse1996}, using a $c$-mBJ parameter of 1.2. This computational scheme demonstrates excellent agreement with experimental bandgap data for \ce{Si}, \ce{Ge}, and $\alpha$-\ce{Sn}, while offering greater efficiency compared to hybrid functional or GW approximation methods~\cite{tran2009, eckhardt2014, polak2017}. Band structures were unfolded into the primitive Brillouin Zone of the diamond cubic lattice using the spectral weight method~\cite{Popescu2010}, as implemented in the \texttt{fold2bloch} code~\cite{Rubel2014}. Relativistic effects, including spin-orbit coupling, were incorporated throughout the calculations, as these are essential to reproduce the experimental band structures of \ce{Ge} and $\alpha$-\ce{Sn}~\cite{eckhardt2014,polak2017}. While a perfectly random alloy is characterized by a \ac{wc-sro} parameter $\alpha = 0$, the actual supercell configurations employed in the bandgap calculations yield a distribution of $\alpha$ values centered near zero. For the 128-atom \ac{sqs} ensemble used in this particular calculation, the reference point is $\alpha = -0.125$, reflecting the practical limitations in generating fully random atomic arrangements in finite-size cells. Computational tractability limits band structure calculations to small supercells (128 atoms), in contrast to the significantly larger supercells ($\approx$ 1780 atoms) employed in the adaptive Bayesian inference approach for \ac{sro} quantification from \ac{exafs} measurements described below. Multiple independent \ac{sqs} configurations (typically 5) are necessary to statistically sample the configurational space and estimate both the mean and standard deviation of the bandgap for the random alloy baseline (error bars in \fref{fig:Fig1}{c} at $\alpha=-0.125$).

\methodTitle{Epitaxial Growth of \ce{Ge}/\GeSn{}{} Core/Shell NWs.} The \ac{vls} growth of \ce{Au}-catalyzed \ce{Ge}/\GeSn{}{} core/shell \glspl{nw} was performed in a \ac{rpcvd} system utilizing ultra-high purity \ce{H2} as the carrier gas, with 10\;\% germane (\ce{GeH4}) and tin(\RNum{4}) chloride (\ce{SnCl4}) as precursor gases. The full experimental procedure has been comprehensively described elsewhere~\cite{meng2016,meng2020}. In brief, \glspl{nw} were synthesized on single-crystal \ce{Ge} (111) substrates by drop-casting \ce{Au} colloids (diameter $\SI{40}{\nm}$, Sigma\hyp{}Aldrich) onto the substrate surface, followed by immediate transfer to the \ac{rpcvd} reactor. The fabrication of the \ce{Ge} \ac{nw} cores was carried out using a carefully optimized two-step thermal process under a constant pressure of $30$ Torr: (i) an initial nucleation phase at $\SI{370}{\degreeCelsius}$ for $\SI{2}{\min}$ to initiate axial nanowire growth at the base, and (ii) subsequent core growth at a reduced temperature of $\SI{300}{\degreeCelsius}$ to promote morphological uniformity by minimizing tapering and kinking. After core synthesis, the \ce{SnCl4} precursor was introduced at $\SI{270}{\degreeCelsius}$ to enable radial overgrowth, yielding coherent and epitaxial \GeSn{}{} shells around the \ce{Ge} cores. For the present study, five samples were fabricated, each featuring a nominal \ce{Sn} incorporation of $(9.5 \pm 0.5)$~at.\%, an average shell thickness of $(142 \pm 12)\;\text{nm}$, and an average wire length of $(4.1 \pm 0.3)\;\mu m$. Significantly, this \ac{nw} platform offers the benefit that the \GeSn{}{} shell remains dislocation\hyp{}free and strain relaxed as a result of growth on the elastically compliant (much smaller radius) \ce{Ge} core, as evidenced by previous studies~\cite{meng2019a,wang2019a,assali2020a}.

\methodTitle{\ac{ald} Growth of \ce{Al2O3} Protective Layer.} A conformal \ce{Al2O3} layer was deposited onto the entire \gls{nw} growth chip via \ac{ald} following a standardized protocol. Immediately prior to \ac{ald} processing, the chip was immersed for $\SI{40}{\second}$ in a 1:1 (v/v) mixture of concentrated 37\;\% \ce{HCl} and deionized (\ce{DI}) water to remove any \ce{Ge}\hyp{} and \ce{Sn}\hyp{}induced native surface oxides. This etching step was succeeded by thorough rinsing with \ce{DI} water and then drying in a nitrogen stream. The cleaned sample was then loaded into the \ac{ald} reactor, which was maintained at a base pressure of $7.5~\text{Pa}$ $(5.6 \times 10^{-2}~\text{Torr})$. Alumina deposition was performed at a substrate temperature of $\SI{200}{\degreeCelsius}$, using trimethylaluminum (TMA) as the aluminum precursor and ozone as the oxidant. Each \ac{ald} cycle consisted sequentially of a $\SI{1}{\second}$ TMA pulse, a $\SI{25}{\second}$ purge, a $\SI{10}{\second}$ ozone pulse, and a final $\SI{30}{\second}$ purge. A total of 24 cycles were executed, resulting in a $(3 \pm 1)~\text{nm}$ \ce{Al2O3} film, determined by spectroscopic ellipsometry on a concurrently processed silicon reference substrate.

\methodTitle{Rapid Thermal Annealing.} RTA was performed using an RTA-610 system (AllWin21 Corp.) to thermally process the \glspl{nw} under a nitrogen (\ce{N2}) environment. Each sample was annealed for \SI{12}{\min} at each temperature from \SIrange{300}{450}{\degreeCelsius}, in $\SI{50}{\degreeCelsius}$ increments. The heating rate was $\SI{100}{\degreeCelsius\per\second}$, followed by a ramp-down at $\SI{50}{\degreeCelsius\per\second}$.

\methodTitle{High-Resolution X-Ray Diffraction of the Annealed Series.} \noindent Structural and compositional characterization of the \ce{Ge}/\GeSn{}{} core-shell nanowire arrays was performed using \ac{hrxrd} on a PANalytical Empyrean system ($\SI{45}{\kilo\volt}$, $\SI{40}{\milli\ampere}$, $1/4^{\circ}$ incident slit, $\SI{2}{\mm}$ mask). Symmetric $\omega$–2$\theta$ scans of the $(333)$ direction (\fref{fig:Fig2}{e}) exhibit well-resolved diffraction peaks from the single-crystalline \ce{Ge}(111) substrate ($\sim90.05^{\circ}$) and the \GeSn{1-x}{x} shell ($\sim88.48^{\circ}$). Quantitative \ce{Sn} content in the \GeSn{}{} shell was extracted using Vegard's law, with a bowing parameter of $0.041~\text{\AA}$~\cite{gencarelli2013}. Gaussian fitting of the \ac{hrxrd} peaks was performed to extract the peak positions and quantify the associated uncertainty. These \GeSn{}{} shell composition estimates, when compared to the TEM-EDX composition maps, show excellent agreement. Analyses across multiple fields of view and \ac{nw} batches confirmed uniform composition and structure for the analyzed samples (Supplementary Information Fig.~S1). Relaxed shell lattice parameters enable a direct correlation of the shell Bragg angle to the \ce{Sn} alloy fraction. For the nanowire geometries and shell thicknesses considered here, prior combined diffraction and modeling studies have shown that \GeSn{}{} shells grown on elastically compliant $\approx\;\SI{50}{\nm}$ diameter \ce{Ge} cores are effectively strain-relaxed~\cite{meng2021a}; under these conditions, the measured out-of-plane lattice parameter can be related to the average \ce{Sn} alloy fraction without requiring additional reciprocal-space mapping. The \ac{hrxrd}-derived compositions reported in \fref{tab:Table1}{} therefore provide a consistent and reliable estimate of the \ce{Sn} content across the annealing series. \ac{hrxrd} composition uncertainty arises from two sources. First, peak-fitting precision is limited by instrumental resolution and counting statistics: $\sigma_a \approx \pm0.003 \AA$. Second, the Vegard-linear calibration carries uncertainty of approximately $\pm$0.3–0.5 at.\% from literature scatter. Error propagation in quadrature yields total composition uncertainty of $\Delta(Sn)\approx\pm0.5$ at.\%.

\methodTitle{\ac{hrtem} Imaging.} \Ac{tem} investigations were performed using a ThemIS 60\hyp{}300 instrument (Thermo Fisher Scientific) operated at an accelerating voltage of $\SI{300}{\kV}$. The microscope was equipped with an image aberration corrector and a Bruker SuperX \ac{edx} detector for spectroscopic analysis. \ac{hrtem} images and selected area electron diffraction patterns were acquired with a Ceta2 CMOS camera, while \ac{stem} images were collected using the \ac{haadf} detector. For plan\hyp{}view analysis, \glspl{nw} were mechanically transferred onto \ac{tem} grids, and cross\hyp{}sectional specimens were prepared via \ac{fib} milling. Insets in \fref{fig:Fig2}{j} display \ac{fft} filtered \ac{hrtem} images. \ac{edx} quantification employed $2\times2$ binning. The parameters for \ac{edx} acquisition were a probe current of $\SI{615}{\pico\ampere}$, a spot size of $\SI{2}{\nm}$, a step size of $\SI{1.5}{\nm}$, and a dwell time of $\SI{8}{\us}$. \ac{edx} uncertainty $(\pm\;0.75\;\text{at.}\%)$ is determined from counting statistics at the measured photon intensity and is consistent with repeated measurements on standard reference samples.

\methodTitle{EXAFS Spectroscopy \& Bayesian SRO Analysis.} \newline
\textit{Measurement:} Fluorescence-yield \ac{exafs} measurements were performed at the \ac{ssrl} beamline 7-3. To eliminate potential substrate background contributions, \ce{Ge}/\ce{GeSn} core/shell \acp{nw} were mechanically exfoliated from their \ce{Ge} (111) growth substrates and transferred onto Kapton tape. Owing to the small volume fraction of \ce{Ge} cores ($\sim$2\%) relative to the \ce{GeSn} shells in the core/shell geometry, the measured \ac{exafs} signal originates almost exclusively from the shell region. Nanowire samples were maintained at liquid helium temperatures ($\SI{7}{\kelvin}$) to suppress thermal disorder. \ce{Ge} K-edge and \ce{Sn} K-edge spectra were collected using a \ac{pips} detector (with \ce{Ga} filter) and a 30-element \ce{Ge} detector (with \ce{Cd} filter), respectively. Data reduction, including dead-time correction, energy calibration, and background subtraction, was performed using \textsc{SIXPACK} and \textsc{ATHENA} software~\cite{ravel2005}. The isolated fine-structure oscillations $\chi(k)$ extraction involved: (i) fitting and subtracting a linear polynomial in the pre-edge region to eliminate instrumental background and non-resonant absorption, and (ii) applying cubic spline functions to model the atomic background in the post-edge region. All data processing steps were performed with the \textsc{ATHENA} software package. Sub-\knn{1} features (\fref{fig:Fig3}{d}) result from the presence of \ce{SnO_x} from surface oxidation of the shells.

\noindent \textit{Bayesian Inference:} Local structural ordering was quantified using \textit{q-SRO}, a custom Bayesian inference framework that extracts short-range order directly from \ac{exafs} spectra. To capture the full coordination environment, \ce{Ge} and \ce{Sn} K-edge spectra were simultaneously fitted in $R$-space ($1.3\;\AA$ to $6.0\;\AA$) using theoretical scattering paths generated from 1728-atom \ac{sro} supercells via \textsc{FEFF7}~\cite{rehr2000}. Critically, scattering paths were \textit{not} averaged: \textsc{FEFF7} calculated individual backscattering amplitudes $f(k)$ and phase shift $\delta(k)$ functions for each unique absorber-scatterer configuration from the \ac{dft}-relaxed supercells. Paths within the same coordination shell and atomic pair type share scattering functions but have occupation weights $N_{\text{eff}}(\alpha)$ determined by the \ac{wc-sro} parameter $\alpha$ (Eq.~S1, Supplementary Information section~2). This preserves full chemical sensitivity without imposing random-alloy assumptions. The fitting model included nine single-scattering paths per edge, accounting for the first three coordination shells (\ce{Ge}$^*$--\ce{Ge}/\ce{Sn}, \ce{Sn}$^*$--\ce{Ge}/\ce{Sn}), plus dominant multiple-scattering paths (e.g., triangular MS \ce{Ge}$^*$--\ce{Sn}$_1$--\ce{Sn}$_2$ and \ce{Sn}$^*$--\ce{Ge}--\ce{Ge}~\cite{shirley2022}). Refined parameters (11 total): (1)~$\alpha$, (2)~$\Delta R_{\ce{Ge-Sn}}$, (3)~$\sigma^2$, (4)~shell-specific coordination numbers (constrained by stoichiometry and supercell \ac{sro}), and (5)~$\Delta E_0$ per edge, with $\Delta R_{\ce{Ge-Sn}}$ and $\sigma_{\ce{Ge-Sn}}^2$ constrained to be identical across both edges for physical consistency. To resolve spectral degeneracies, we implemented Bayesian inference with \ac{dft}-calculated bandgap priors (Supplementary Information section~6):
\begin{equation*}
P(\alpha|\mathcal{D}_{\text{EXAFS}}) \propto \mathcal{L}(\mathcal{D}_{\text{EXAFS}}|\alpha) \cdot P_{\text{DFT}}(\alpha),
\end{equation*}
where $\mathcal{L}(\mathcal{D}_{\text{EXAFS}}|\alpha)$ represents the \ac{exafs} likelihood function derived from spectral goodness-of-fit, and $P_{\text{DFT}}(\alpha)$ is the informative prior defined by the consistency between the theoretical bandgap of the short-range order structure and experimental optical absorption data. Final spectra were ensemble averaged over 180 \ce{Sn} and 50 \ce{Ge} absorber sites per supercell to account for local structural heterogeneity. This \ac{dft}-informed prior approach reduces \ac{sro} parameter uncertainty by approximately 50\% compared to uninformed fitting (uniform prior) while maintaining unbiased central estimates (Supplementary Information Fig.~S10 and \fref{tab:Table1}{}), demonstrating that the framework enhances statistical precision without introducing systematic bias. The \ac{msrd} for each scattering path and coordination shell was computed independently using the \ac{dft}-relaxed supercell geometries (Supplementary Information Eq.~S2). Complete details of the Bayesian inference algorithm, clustering methodology, and validation against uninformed priors are provided in Supplementary Information sections~2 and~6.

\methodTitle{Optical Characterization.} 
Quasi-continuous wave (qCW) \ac{pl} measurements were performed with samples mounted in a liquid-nitrogen-cooled cryostat featuring a \ce{CaF2} optical window. Sample excitation was provided by a $\SI{1}{\watt}$, $\SI{808}{\nm}$ diode laser, which was electrically modulated as a square wave at $\SI{10}{\kHz}$ with a 50\;\% duty cycle. The emitted \ac{pl} was collected by a reflective parabolic mirror and directed into a Bruker Invenio-R Fourier-transform infrared (FTIR) spectrometer for spectrally resolved analysis. The spectrometer was operated in step-scan mode, and the signal was detected using a liquid-nitrogen-cooled \ac{mct} detector coupled to a lock-in amplifier to reject thermal background. For power-dependent measurements, the incident excitation density was varied using a series of neutral density filters. The laser spot area on the sample surface was determined to be $\SI{3670}{\micro\m\squared}$ via the knife-edge method, corresponding to an excitation density range of approximately $\SIrange{1}{22}{\kilo\watt\per\cm\squared}$. Spectral line shapes were fit using a model based on the convolution of a Gaussian function with a \ac{jdos} term~\cite{Assali2021a, vina1984c}, described by:
\begin{multline} 
    I_{PL} = A \left[ \sqrt{E - E_g} \times \exp \left( \frac{-E}{k_B T} \right) \right]
    \ast \\ \left[ \frac{1}{\gamma \sqrt{2\pi}} \exp\left( \frac{-E^2}{2 \gamma^2} \right) \right],
\label{eq:jdos_lineshape}
\end{multline}
where $A$ is a scaling constant with units of (energy)$^{1/2}$, $E_g$ is the bandgap energy, $T$ is the absolute temperature, and $\gamma$ is the spectral broadening parameter. The first term, representing the \ac{jdos} multiplied by the Boltzmann distribution, models the emission profile for parabolic bands under low-excitation conditions. The second term accounts for broadening effects—including alloy composition fluctuations and strain inhomogeneities—through convolution with a Gaussian function~\cite{schubert1984,ouadjaout1990}. More details related to the optical analysis are presented in Supplementary Information section~3.

\bigskip \medskip
\paperSection{CODE AVAILABILITY}
All codes that support the findings of this study are available from the corresponding authors upon reasonable request.

\bigskip \medskip
\paperSection{ACKNOWLEDGEMENTS}
This work was supported by $\mu$-ATOMS, an Energy Frontier Research Center funded by the U.S. Department of Energy (DOE), Office of Science, Basic Energy Sciences, under award $\text{DE\hyp{}SC0023412}$ and SubAward No. $\text{UA2023\hyp{}351}$. This study includes work supported by the DOE Office of Science, Office of Workforce Development for Teachers and Scientists, through the Office of Graduate Student Research (SCGSR) program, administered by the Oak Ridge Institute for Science and Education under contract $\text{DE\hyp{}SC0014664}$. Additional support was provided by the U.S. National Science Foundation under grant $\text{DMR\hyp{}2003266}$ and by the National Institute of Standards and Technology (NIST). We acknowledge the use of the Stanford Synchrotron Radiation Lightsource (SSRL), SLAC National Accelerator Laboratory, supported by the DOE Office of Science, Office of Basic Energy Sciences, under contract $\text{DE\hyp{}AC02\hyp{}76SF00515}$. This research used resources of the National Energy Research Scientific Computing Center (NERSC), a DOE Office of Science User Facility, under contract $\text{DE\hyp{}AC02\hyp{}05CH11231}$ and NERSC Award $\text{BES\hyp{}ERCAP0027056}$. Additional computing support was provided by the Sherlock cluster at Stanford University; we thank Stanford University and the Stanford Research Computing Center for computational resources and support. The authors also acknowledge the George Washington University High Performance Computing facility for computing support. Part of this work was performed at the Stanford Nano Shared Facilities (SNSF), supported by the National Science Foundation under award $\text{ECCS\hyp{}2026822}$. Reference to specific software packages, equipment, and materials is only for completeness and does not represent an endorsement by NIST or the Federal Government of the USA.

\bigskip \medskip
\paperSection{AUTHORS INFORMATION}
A.A. carried out thermal annealing and \ac{exafs}-related theoretical analysis and measurements. J.Z.L. carried out the epitaxial growth of the \ce{Ge}/\GeSn{}{} \glspl{nw}. J.M. and K.M. provided the PL measurements. J.C.W supported the EXAFS analysis. L.V. and A.M.M. performed all the TEM characterizations of the annealed \glspl{nw}. S.C. and T.L. performed the \ac{dft} calculations and conducted the large-scale \ac{exafs} simulations using supercells generated from machine-learning atomistic simulations. P.C.M. conceptualized the research, acquired funding for it, and led the research. All authors contributed to the manuscript. 
Corresponding Authors:\\
\textcolor{blue}{$^{*}$}\href{mailto:anatt@stanford.edu}{anatt@stanford.edu}\\\textcolor{blue}{$^{\dagger}$} \href{mailto:pcm@slac.stanford.edu}{pcm@slac.stanford.edu}

\bigskip \medskip
\paperSection{NOTES}
The authors declare no competing financial interests.

\let\oldaddcontentsline\addcontentsline
\renewcommand{\addcontentsline}[3]{}

\bibliography{references_combined}
\bibliographystyle{naturemag}

\end{document}